\documentclass[nohyper,12pt,letterpaper]{JHEP}
\usepackage{epsfig}
\newfont{\frak}{eufm10 scaled 1200}

\newfont{\Bbb}{msbm10 scaled 1200}     %instead of eusb10
\newcommand{\mathbb}[1]{\mbox{\Bbb #1}}
\DeclareSymbolFont{AMSa}{U}{msa}{m}{n}
\DeclareSymbolFont{AMSb}{U}{msb}{m}{n}
\let\Box\relax
\DeclareMathSymbol{\Box}{\mathord}{AMSa}{"03}

\def \t{\theta}
\def \b{\beta}
\def \L{\Lambda}
\def \f{\phi}
\def \r{\rho}
\def \p{\pi}
\def \P{\Pi}
\def \gr{\nabla}

\def \T{\tau}
\title{Evidence for Winding States in Noncommutative Quantum Field Theory}

\author{W. Fischler, E. Gorbatov, A. Kashani-Poor, S.
Paban, P. Pouliot\\
  Department of Physics\\
  University of Texas, Austin, TX 78712\\
E-mail: \email{fischler, elie, kashani, \\
\hspace{1.2cm}paban, pouliot@physics.utexas.edu}}

\author{Joaquim Gomis \\
Departament ECM, Facultat de Fisica \\
Universitat de Barcelona \\
and \\
Institut de Fisica d'Altes Energies \\
Diagonal 647, E-08028 Barcelona, Spain \\
E-mail: gomis@ecm.ub.es}

\abstract{
We study noncommutative field theories at finite temperature to learn more
about the
degrees of freedom in the non-planar sector of these systems. We find
evidence for
winding states. At temperatures for which the thermal wavelength is \mbox{smaller}
than the
noncommutativity scale, there is a drastic reduction of the degrees of freedom
in the non-planar sector. In this regime, the non-planar sector has
thermodynamics
resembling that of a
$1+1$ dimensional field theory. }


\preprint{\hepth{0002067}\\ UTTG-03-00 \\ UB-ECM-PF-00/02}
\begin{document}

%%%%%%%%%%%%%%%%%%%%%%%%%%%%%%%%%%%%%%%%%%%%%%%%%%%%%%%%%%%%%%%%%%%%%%%%%%%%
%          Table of contents automatic !!!                                 %
%%%%%%%%%%%%%%%%%%%%%%%%%%%%%%%%%%%%%%%%%%%%%%%%%%%%%%%%%%%%%%%%%%%%%%%%%%%%

\section{Introduction}
% ==========================================================================

Gauge theories on noncommutative spaces can be obtained by taking the
infinite tension limit of string theory in
the presence of a very strong B-field \cite{Connes:1998cr,Seiberg:1999vs,
Minwalla:1999px,Bigatti:1999iz}. One can
thus look upon these field theories as
interesting limits of string theory in such an extreme environment.
Regardless of their ancestry, these field theories have very
interesting properties \cite{Filk:1996dm}--\cite{Grosse:2000yy},
making their study fascinating in its own right. Many unusual properties already
emerge in scalar field theories. In
particular, one sees the emergence of unexpected infrared behavior in
correlation
functions \cite {Minwalla:1999px} (we will refer to this paper in the following
as MSV) unlike anything seen in conventional field theories. These long
distance
correlations appear in massive theories and are not related to any massless
fields in the Lagrangian. Rather, these infrared effects are due to the Moyal
phases. As can be seen in Feynman diagrams, it is the integration over the high
ultraviolet modes that is responsible for this peculiar infrared behavior.
The UV
and IR are intertwined in a manner never seen in local
field theory on commutative spaces. MSV have a
proposal for what kind of degrees of freedom are responsible for these IR correlations: their claim is that in the
zero slope limit and large B field, the
closed string states do not decouple.

To gain some more insight into the structure of these noncommutative
field theories, we subject them to a heat bath and consider various
thermodynamic
quantities. We will show that the partition functions of these theories get
contributions at high temperature from configurations winding around the
temporal
direction. This implies that there are extended objects in these theories
that have
the ability to wind, which seems consistent with the suggestions of MSV. We
also test
this picture by compactifying the spatial commuting direction to a circle and
again find contributions to the partition function, this time from
configurations
winding around the spatial circle.

It is useful to consider a simple field theory like
massive $\phi^4$ and the supersymmetric Wess-Zumino model. We were motivated to
include a supersymmetric example in order to show that the existence of
winding states persists in theories where the sensitivity to the UV physics is
softer. Because of the special UV properties of the supersymmetric case, the IR
sensitivity is milder and allows us to limit ourselves to the lowest order
in perturbation theory when calculating thermodynamic quantities.

The paper is organized as follows: in
section 2 we will describe in some detail the thermodynamics of noncommutative $\phi^4$. There we present the
calculation of the free energy to $O(g^2)$ and discuss the various
divergences and also show how the
winding states appear. In section 3, we investigate whether supersymmetric
systems
whose UV behavior are less singular affect the existence of winding states. For
this purpose, we consider the Wess-Zumino model and find again the existence of
winding states. We
calculate the free energy, from which the entropy, internal energy
and specific heat can be derived. In section 4, we return to the
$\f^4$ theory and show how to calculate the free energy in the ``mean field"
approximation. In section 5, we discuss the implications of these results and
briefly repeat the calculations at zero and finite temperature for the case
where the sole commuting direction is a
circle. We show that at finite temperature, there are no membrane
states wrapping around the torus spanned by the temporal and $x_3$
cycles. Instead, we find winding states winding each of the
cycles independently.

We then conclude with some speculations.

\section{The thermodynamics of noncommutative $g^2\phi^4$ in perturbation
theory}

We present a perturbative calculation of the free energy, F,  which assumes
that the coupling constant
$g^2$ is small. As
we will see, the appearance of IR divergences invalidates the perturbative
expansion. Nevertheless, we believe it is instructive to go through this
exercise as
 it already shows the existence of winding states. Later in section 4 we
will have to
use a mean field approximation to ensure that long range correlations are
properly
 screened.

In this paper, we will restrict ourselves
to the case where the noncommutativity is solely among the spatial
directions.. Without loss of generality then in this 4-dimensional
case, we can choose the (1,2) plane to be noncommutative and leave time
and the third spatial dimension
commutative.

At finite temperature, the Feynman rules are the ones that correspond to a
theory
with ``time" being a circle of radius $\b$, where $\b$ is the inverse
temperature, $\b = 1/T$.
This implies that the integrals over frequencies in diagrams are replaced by
discrete
sums over the so-called Matsubara frequencies. The noncommutativity in
the (1,2) plane manifests itself in
Feynman diagrams through the appearance of Moyal phases in vertices. These
Moyal phases depend on momenta in the
(1,2) directions.

When evaluating the contributions to the free energy, one needs to consider
both planar and non-planar
diagrams. The planar contributions are very much like in the commutative
theory except for some combinatoric
factors \cite{Filk:1996dm}. Inspite of the changed combinatorics, the
behavior in
the UV is still controlled by zero temperature physics. In order to
illustrate the
divergences that occur in the perturbative expansion of the free energy in the
noncommutative theory, we will first concentrate on the non-planar
contributions to
lowest order in
$g^2$. This will also enable us to show the appearance of
contributions from winding
states to the free energy.

 The leading non-planar contribution at order $g^2$ to the free
energy comes from a two-loop diagram.

\vskip1cm
\epsfig{file= 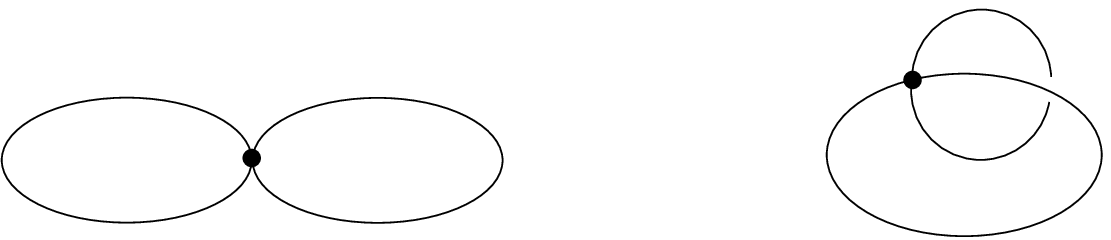} 

\centerline{ {\bf Fig.1:} Planar and nonplanar two loop contributions to the
free energy.}

\vskip1cm

 This diagram is
divergent, where some of the divergences can be
interpreted as originating in the UV and some divergence
 can be attributed to the IR. The UV divergences will be shown to arise from
zero temperature
physics and similar UV divergences will appear in higher order terms in the
expansion in $g^2$. Also, in a
manner similar to
 the IR divergences that appears at lowest order, each successive order of
perturbation theory will be plagued
by
 IR divergences. These IR divergences grow more singular with increasing
order in the perturbative
expansion, invalidating the expansion.
We will return to the formal application of this mean field approximation
in section
4 after the study of the
$O(g^2)$ contribution to the free energy in the Wess-Zumino model.

 The contribution to the free energy from this two-loop non-planar diagram
is:

\begin{equation}
-{g^2} T^2 \sum_{n=-\infty}^\infty\ \sum_{l=-\infty}^\infty \int \frac{d^3p}{(2
\pi)^3}\,
\frac{d^3k}{(2
\pi)^3}
\,\frac{e^{i\, p\,  \t \, k}} {(\frac{4\pi^2 n^2}{\b^2} + p^2 +
M^2)(\frac{4\pi^2
l^2}{\b^2} + k^2 + M^2)},   \label{eight}
\end{equation}
where $p \t k = \t (p_1k_2 - p_2k_1)$ and $p^2 =
{p_1}^2 + {p_2}^2 + {p_3}^2$.

We will be interested in temperatures much larger than the mass, so that to
leading order in $\b M$, it can be neglected.

The integration over the momenta can be done by introducing Schwinger
parameters. In order to show how
contributions of winding configurations appear, we will look at equation
(\ref{eight}) in some detail.
The integral appearing in this expression can after some
manipulations be rewritten as:

\begin{equation}
-{g^2} T \sum_{n,l} \int \frac{d^3p}{(2 \pi)^3} \,\frac{1}{(\frac{4\pi^2
n^2}{\b^2} + p^2)(4\pi^2 (l^2\b^2 +|\t p|^2) + \L^{-2})}\,, \label{poisson}
\end{equation}
where $|{\t} p| = \t({p_1}^2 +
{p_2}^2)^{1/2}$ and $\Lambda$ is an ultraviolet cutoff similar to the one used
by MSV. We also have neglected
the mass in obtaining this expression since we are working in the regime
$\b M \ll 1$.

The sums over $n$ and $l$ can be performed and will lead to the following
formula involving Bose distributions

\begin{eqnarray}
-{g^2}T \int \frac{d^3p}{(2 \pi)^3} \, \frac{1 + 2
n_{\b}(|p|)}{2|p|}\frac{1 + 2 n_{1/\b}(2\pi|\t p|)}{4\pi|\t p|},  \label{uv}
\end{eqnarray}
where  $|p|= ({p_1}^2 + {p_2}^2 + {p_3}^2)^{1/2}$ and $ n_{\b}(|p|) =
\frac{1}{e^{\b
|p|} - 1}$ is a Bose distribution at temperature
$T = 1/\b$.

Note the appearance in the last expression of two Bose distributions, one at
temperature $T$ and the other at
temperature $1/T$. This would be a rather bizarre state of affair for a system
in thermal equilibrium. One possibility
is that the system is not in equilibrium, but why then in this very special
way, with one set of the degrees of
freedom distributed thermally at temperature $T$ and the other at temperature
$1/T$?
We will choose the option that the system is indeed in equilibrium, and
that there are configurations, if one thinks
in a path integral approach to the partition function, that wind around the
temporal circle. If one looks at the
integral appearing in equation (\ref{uv}), one can see the contributions of
winding
states with momentum $|{\t}p|$ to the free
energy.

Another interesting feature of equation (\ref{uv}) is that the $p_3 = 0$
sector is
invariant under the interchange of $\b$ and $2 \p\, \t/{ \b}$. This
transformation
looks like T-duality where the string scale ${l_s}^2$ is replaced by
$\t$. We should remind the reader though that this property was obtained in
the
$\b M \ll 1$ limit. We do not know whether this duality invariance persists
beyond this order to the full theory.

The expression in equation (\ref{uv}) has UV divergences and IR
divergences. The IR
divergence is linear and comes from
the small
$(p_1,p_2)$ region of integration. This divergence, as will be shown
later, is cured by the mean field
approximation.

The UV divergent terms in the non-planar sector are calculated in the
appendix. Apart
from a zero temperature contribution to the vacuum energy,
there is a linearly divergent term which is also linear in the temperature.
This
is a rather unusual contribution to the free energy. Indeed, this term does not
contribute to the internal energy of the system but does provide a temperature
independent contribution to the entropy. This contribution to the entropy is
due to the degeneracy of the winding
states. As can be seen in the Bose distribution, their dispersion relation
does not involve the momentum along the commuting spatial dimension, $p_3$. The
very large number of winding states with momenta
$p_{1,2} = 0$ are the source of the linear divergence. This divergence
disappears
when screening of the large distance correlations is taken into account. The
supersymmetric case does not have such a divergence at the leading order in
$g^2$
because of its milder behavior in the IR.

\section{The thermodynamics of the noncommutative Wess-Zumino model}

The Lagrangian for the noncommutative Wess-Zumino model is:

$${\cal L} = i \partial_\mu \bar{\psi}\bar{\sigma}^\mu \psi + A^* \Box\, A
-\frac{1}{2} M
\psi
\psi -\frac{1}{2}M \bar{\psi}\bar{\psi} -g \psi * \psi\,  A -
g\bar{\psi}*\bar{\psi}\, A^* -F^*  F \,,$$
where $F$ is given by
$$F= - M A^* - g A^* * A^* \,.$$

The calculation of the free energy to $O(g^2)$ proceeds along the same
lines as in
the bosonic case. In this case there are additional contributions due to new
interactions among the bosons and bosons with fermions. The fermions, as
usual, have
frequencies that are odd multiples of the temperature.

Three pairs of diagrams contribute at order $g^2$, each pair consisting of
a planar
and a non-planar piece. They sum to give

\begin{eqnarray}
F/V &=& -g^2 \int \frac{d^3 p}{(2\p)^3} \frac{d^3 k}{(2\p)^3}
\frac{1+e^{i\, p\, \t
\, k}}{
\, \omega_p
\, \omega_k} \Big(n_B (\omega_p) + n_F(\omega_p)\Big)\, \Big(n_B(\omega_k) +
n_F(\omega_k)\Big)  \nonumber \\  &+& g^2 T^4 {\cal O}(\frac{M^2}{T^2})      \label{wzfe}
\end{eqnarray}
% $$+M^2{g^2} T^2 \sum_{n=-\infty}^\infty\ \sum_{l=-\infty}^\infty \int
% \frac{d^3p}{(2
% \pi)^3}\,
% \frac{d^3k} {(2
% \pi)^3} ({1+ e^{i\, p\,  \t \, k}}) G(M,\b,p,k)\,,  \label{wzfe}
% \end{equation}
with $\omega_p =\sqrt{p^2 + M^2}$ and $n_{B,F}(\omega_p) = {1}/({e^{\b\,
\omega_p}
\, \mp\, 1})$.
% \begin{equation}
% G(M,\b,p,k) = \left( \frac{1} {(\frac{4\pi^2 n^2}{\b^2} + p^2
% + M^2)(\frac{4\pi^2 l^2}{\b^2} + k^2 + M^2)(\frac{4\pi^2 (n+l)^2}{\b^2} +
% (p+k)^2
% + M^2)}\right)
% $$
% $$- \left( \frac{1} {(\frac{\pi^2 (2 n+1)^2}{\b^2} + p^2
% + M^2)(\frac{\pi^2 (2l+1)^2}{\b^2} + k^2 + M^2)(\frac{4\pi^2
% (n+l+1)^2}{\b^2} +
% (p+k)^2 + M^2)}\right).
% \end{equation}
Notice that this expression for the free energy at high temperature ($\b
M\ll1$) is ``T- duality invariant" as in the bosonic case. This property applies to
the sector with zero momentum along the commuting direction.
This result can be seen by expanding one of the Bose distributions in a
series of
exponentials,

$$\frac{1}{e^{\b (p^2 + M^2)^{1/2}} -1} = \sum_{n=1}^{\infty}e^{-\b\, n\,
(p^2 +
M^2)^{1/2}}.$$
Then performing the integral over $d^3p$, neglecting the mass,
produces the series,

$$ \sum_{n=1}^{\infty} \frac{1}{n^2\b^2 + |\t k|^2}. $$
A similar expression can be obtained for the integration over the Fermi
distributions, except that in the fermion case the series is an alternating
one. The
presence of winding states is then clear as well as the ``T- duality
invariance" of
the free energy.

\vskip1cm
\epsfig{file= 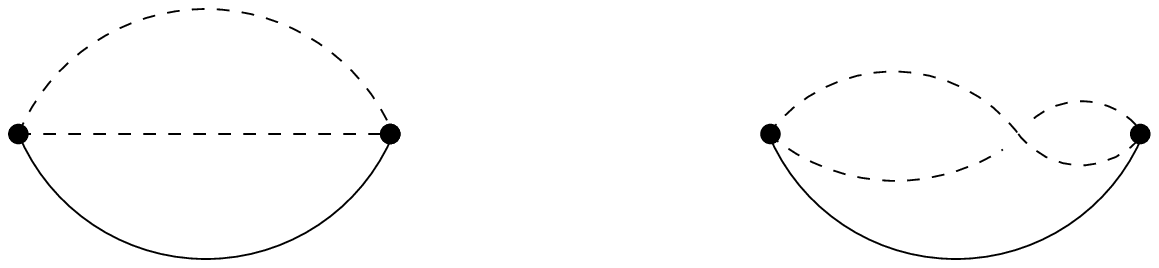} 

\centerline{ {\bf Fig.2:} Planar and non-planar two loop fermionic contributions to the
free energy.}

\vskip1cm
\epsfig{file= 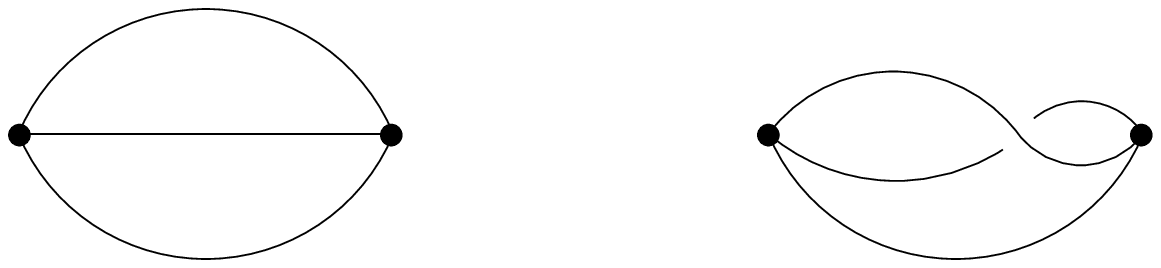} 

\centerline{ {\bf Fig.3:} Planar and non-planar two loop bosonic contributions to the
free energy.}

\vskip1cm

The expression (\ref{wzfe}) for the non-planar contribution to the free
energy can be
further evaluated to yield

\begin{equation}
-\frac{(2\p)^3}{\sqrt{2}}\frac{g^2}{\t \b^2} \int_0^{\p} d \eta\,
\int_0^\infty dq \,
\frac{\tanh{(\frac{\p sin(\eta)}{2}\frac{\t q}{\b^2})}}{\sinh q}.
\end{equation}

At high temperature, $\b M \ll 1$, we consider two limiting cases, $T^2\ll 1/\t
$ and $T^2 \gg 1/\t. $ \footnote{We also assume that $\t M^2 \ll 1$, i.e. the
Compton wavelength of the scalars is larger than the length scale
associated to the
noncommutativity.} The first limit corresponds to the case where the thermal
wavelength is bigger than the length scale associated to the
noncommutativity. In
this limit, one might expect to recover conventional high temperature behavior.
Indeed, the leading non-planar contribution to the free energy here is

\begin{equation}
\frac{F}{V} \Big|_{\rm np} \sim - g^2 T^4 \,.
\end{equation}

In the other limit, $T^2 \gg 1/\t$, the thermal wavelength is smaller than the
noncommutativity scale. We therefore expect novel behavior. We find the
free energy
in this regime to be

\begin{equation}
\frac{F}{V} \Big|_{\rm np} \sim  -g^2 \frac{T^2}{\t} \log{T^2 \t} \,,  \label{red}
\end{equation}
where the logarithm was obtained by numerical methods.
In this case, the contribution to the free energy resembles the contribution of
a massless gas in 1+1 dimension, modified by a logarithm.

One is tempted to conclude from these results that there is a drastic
reduction of
degrees of freedom in the non-planar sector! For thermal wavelengths larger
than the
noncommutativity scale, the system behaves as a relativitic $3+1$
dimensional gas.
For wavelengths smaller than this scale, the system behaves roughly like a
$1+1$
dimensional gas. As the temperature is raised past
$ 1/\t^{1/2}$ it looks like the winding states do behave as strings. Up to the
logarithm, one is reminded by the temperature dependence in equation
(\ref{red}) of
the result obtained by Atick and Witten
\cite{{Atick:1988si}} for the high temperature behavior of strings.
We do not know whether this is just
a coincidence or whether there is some deeper significance to this
behavior. A test
of this high temperature dependence of the free energy due to the non planar
sector would be to study higher dimensional examples where there are more
noncommuting spatial dimensions. It is amusing to note that the non-planar
sector of
this system exhibits, to leading order at very high temperature, the
equation of
state $p = \rho$, $p$ being pressure and $\rho$ energy density.

\section{The thermodynamics of noncommutative $g^2\phi^4$ in the ``mean
field" approximation}

We now go back to the noncommutative $g^2\phi^4$ theory of section 2.
There, the appearance of IR divergences ruins the validity of the perturbative
expansion (in $g^2$) for the free energy. This
is a familiar phenomenon in thermal physics. For example, in the case of a
plasma of electric charges in QED, one finds IR divergences in
perturbation theory for the free energy. This can be traced to the large
distance behavior of the correlation function, $\langle\r(x)\r(0)\rangle.$
This IR
divergence disappears after summing a series of diagrams contributing to the
correlation function between charge densities,
$\langle\r(x)\r(0)\rangle.$ This series of diagrams is a geometric series
in the
one-particle irreducible charge density correlation function. When this series
is summed, one finds that the charge densities are screened. The
screening mass or Debye mass is the IPI charge density correlation function
 evaluated at zero momentum.

In the case of noncommutative $\f^4$, the IR divergence already appears at
zero temperature and is again screened by summing a
geometric series in the one-particle irreducible two-point function (MSV).

In the
finite temperature case, we will approach the problem of IR divergences by
treating
the gas in a mean field approximation. In this approximation, each quantum
of the
scalar field
 is moving in the average field produced by the other scalars. Unlike more
traditional cases where the mean field approximation is used, this case has the
particular feature that the mean field will depend on the momentum of the
particle that moves in this background. This momentum lies in the
noncommutative plane. Using this approximation, we will show how the long
range correlations are screened.

The Hamiltonian H is:

$$ H = \frac{1}{2}\int d^3x\,(\P^2(x) + (\gr\f(x))^2 + M^2\f^2(x)) +
g^2\int d^3x\, \f(x)*\f(x)*\f(x)*\f(x).$$

The free energy F is obtained from the logarithm of the partition function
$Z$, where $Z = {\rm tr}\, e^{-\b H}$, as $-\b F = \log Z$. When
$g^2$ is small, it can be formally written as
$F_0$, the free gas contribution to F, corrected by $O(g^2)$ contributions. We
should remind the reader that this procedure fails because of IR
divergences. We present this expression for F here because it will be useful
when we get to the mean field approximation:

$$ \frac{\b F}{V} = \int \frac{d^3p}{(2 \pi)^3} [\frac{\b}{2} (p^2 +
M^2)^{1/2} +
\log(1 -
\exp(-\b (p^2 + M^2)^{1/2}] + O(g^2).$$
The contribution from the free gas to the previous formula could have
been obtained by calculating a one-loop diagram in the path integral
formulation.

In the mean field approximation, a picture emerges where the system looks
like a
free gas for which the dispersion formula for the energy of a quantum is
modified by the average effect of the other quanta.

Formally, this will be
implemented in the path integral by using the following action:
\begin{eqnarray}
S = \int_0^\b d\tau \int \frac{d^3k}{(2 \pi)^3} \left[(\partial_t\f)^2(k) -
k^2\f^2(k) - M^2\f^2(k) - g^2\langle \f^2 \rangle_{\t k} \f^2 \left( k \right)
\right].
\label{action}
\end{eqnarray}
where $\langle\f^2\rangle_{\t k}$ satisfies an integral equation:

$$\langle\f^2\rangle_{\t k}=1/\b \sum_n \int \frac{d^3p}{(2 \pi)^3}
\frac{2+e^{i\, p \,\t \, k}}{\frac{4\p^2 n^2}{\b^2} +p^2 +M^2
+g^2\langle\f^2\rangle_{\t p}}. $$  We will not solve this integral
equation exactly
but rather find the solution to it
 as a power series expansion in $g^2$. In this paper we will confine ourselves
to work with the solution to leading order.

In the limit of high temperature, $M \b \ll 1$, we neglect
to first order the mass M and we sum over $n$, and obtain to leading order
$\langle\f^2\rangle_{\t k}$.

\begin{eqnarray}
\langle\f^2\rangle_{\t k} &=& \sum_n \frac{1}{4\p^2n^2\b^2 + (2\p|\t k|)^2
+\L^{-2}}
\\ &=& \frac{1}{2\b\sqrt{(2\pi|\t k|)^2+\L^{-2}}} (1 + 2 n_{1/\b}(\sqrt{4\pi^2|\t
k|^2+\L^{-2}}) \,.    \label{phi2}
\end{eqnarray}

Notice that the Bose distribution in the expression for $\langle\f^2\rangle$
is unusual as it distributes $k_{1,2}$ according to a ``temperature" $\T $,
where  $\T = 1/\t T $, i.e. proportional to the inverse temperature. One
can interpret this property, as was argued earlier, to be evidence for
winding states. Indeed, the summation that appears in $\langle\f^2\rangle$
is over
an integer multiple of the length of the temporal circle rather than the
inverse
radius as would be the case for momenta.
The free energy can
then be written as a ``tree" contribution plus a one-loop contribution.

\begin{equation}
\frac{\b F}{V} = M^2 \langle\int d^4x\, \f^2(x)\rangle+\frac{\b g^2}{V}
\langle\int d^4x\,\f(x)*\f(x)*\f(x)*\f(x)\rangle+$$
   $$\int \frac{d^3p}{(2 \pi)^3} {\left[\frac{\b}{2} \sqrt{p^2
+M^2+g^2\langle\f^2\rangle_{\t p}}
 +\log\left(1 - \exp(-\b {(p^2 +M^2+g^2\langle\f^2\rangle_{\t p})}^{1/2} \right)
\right] }.
\label{ir}
\end{equation}
In this last equation, the average value of

$$   M^2 \int d^4x\, \f^2(x)
+\frac{\b g^2}{V}
\int d^4x\,\f(x)*\f(x)*\f(x)*\f(x)$$
is evaluated by using the action S,
equation (\ref{action}), in a path integral calculation.

As expected and demonstrated explicitly in the appendix, the IR divergences
disappear.

The one-loop contribution

$$\int \frac{d^3p}{(2 \pi)^3} \left[ \frac{\b}{2} \sqrt{p^2
+M^2 + g^2 \langle \f^2 \rangle_{\t p}}
 +\log{ \left( 1 - \exp{(-\b {(p^2 + M^2 + g^2 \langle \f^2 \rangle_{\t
p} )}^{1/2}} \right) } \right] $$ can be visualized as a single quantum of the
scalar field  propagating around a loop with a momentum that at $O(g^0)$ is
thermally
distributed at temperature $T$. This particle is moving in the presence of a
background, $\langle\f^2\rangle$, which it interacts with at $O(g^2)$ and
that consists of
states winding around the temporal circle of radius $1/T$. These winding
states have momenta solely in the noncommutative plane. It is quite tempting
to think of them as closed strings reminiscent of the suggestions made in MSV.

To be complete, one should study the UV divergences and see whether they are
associated to zero temperature effects. We will not attempt this study here. It
should be kept in mind that it is not completely clear yet whether this
theory is renormalizable, even though some evidence exists.

\section{Winding states around spatial cycles}

The presence of winding states, which in the finite temperature case wind
around the temporal circle, can also be noticed when one compactifies the
commuting spatial dimension, $x_3$. The states winding around $x_3$ can be
detected for example if we calculate the free energy for this case.

The calculation of the previous section can be repeated, with each $\int
\frac{dp_3}{2\p}$ replaced by $\frac{1}{2\p L}\sum_l$, where l is an integer.

To show the existence of winding states, it is sufficient to again consider the
integral equation for the two point function:

$$\langle\f^2\rangle_{\t k}= \frac{1}{\b L} \sum_{n,l} \int
\frac{d^2p}{(2\p)^2}
\frac{e^{ip \t k}}{\frac{4\p^2n^2}{\b^2} +\frac{4\p^2l^2}{L^2}+p^2 +M^2
+g^2\langle\f^2\rangle_{\t p}} \,.$$

In the limit of high temperature, $\b M \ll 1$, and small radius, $LM \ll
1$, we
find for $\langle\f^2\rangle_{\t k}$, upon performing the $\int d^2p$,

$$\langle\f^2\rangle_{\t k} = \sum_{n,l}\frac{1}{4\p^2 n^2 \b^2 + 4\p^2 l^2 L^2
+(2\p|\t k|)^2 +\L^{-2}} \,.$$

This last expression clearly shows the presence of winding states around
both circles. Again these states become prevalent in the small radii
limit, which is to be expected since in these limits these states are very
``light".

This formula for $\langle\f^2\rangle_{\t k}$ also suggests that there are
{\em no}
wrapping states around the 2-torus spanning the temporal and $x_3$ dimensions,
which would look like an additional  piece $(n\, l\, \b \, L)^2$ in the
denominator
of this equation.

The same result holds in the supersymmetric case.
In some ways, this could have been expected. Imagine subjecting this system
to a thermal environment. In this case, the bosons and fermions
have different distributions and  we should not therefore expect a
drastically different behavior from the non-supersymmetric cases.

One might speculate about the existence of multidimensional extended objects in
higher dimensional theories where several $\t_{ij}\not= 0$. To test this, we
briefly consider the example of the six-dimensional $\f ^3$
theory, with various $\t_{ij}\not= 0$.

At high temperature and after compactifying the
commuting $x_5$ on a small circle, it should be possible to see whether or not
wrapping states emerge. This can be tested, as we learned above, simply by
calculating
$\langle\f^2\rangle_{\t p,\t'p'}$. We obtain

$$\langle\f^2\rangle_{\t p,\t'p'} =\sum_{l,n}\frac{1}{(- 4\p^2 p_i \theta^{i j}
\theta_{j k} p^k +4\p^2\b^2 n^2+4\p^2 L^2 l^2)^2} \,.$$

We thus conclude that there are no wrapping states, just winding states.

\section{Conclusions}
What is the physics that is responsible for the existence of these winding
states?

Clearly the UV-IR connection that exists in the noncommutative plane is
crucial. The expressions which reveal the existence of the winding
states show that this peculiar IR behavior originates in the UV
region of integration. One might then conclude that
the uncertainty relationship between $x_1$ and $x_2$ in the case where $\t
_{12}\not= 0$ is implemented by or implies the existence of one-dimensional
structures. In other words, measuring $x_1$ sharply renders the other
direction
$x_2$ totally uncertain and maybe this is intimately related to the
existence of extended one-dimensional objects in a rather mysterious way.
These ``strings" can then wind around various cycles.
However, we did not find multidimensional objects wrapping multicycles in
higher
dimension.

We have discovered from studying the high temperature regime of the
supersymmetric
theory the existence of two different behaviors. We believe that these
results are
valid in general, but we have only been able to calculate reliably to
leading order
in the supersymmetric case. When the thermal wavelength is bigger than the
noncommutativity length scale, we find that the non-planar sector behaves as a
conventional $3+1$ dimensional relativistic gas. For the thermal wavelength
smaller than
the noncommutativity scale, we observe a reduction of the degrees of
freedom in the
non-planar sector. The non-planar sector now behaves as a $1+1$ (not, as might
have been expected, $2+1$) dimensional gas. This seems to indicate a drastic
reduction of the degrees of freedom in that sector. Equivalently, we find
the equation of state of this sector is approximately that of an
incompressible gas
with equation of state $p=\rho$.

It would be interesting to see whether this behavior persists in higher
dimensional
theories.

In the theories we have studied, the fascinating behavior of the non-planar sector is overshadowed by the planar sector. Thus, the
challenging question remains of what theory captures exclusively the
physics of the
non-planar sector.

%=========================================================================
\acknowledgments
%=========================================================================

The work of WF, EG, JG, AK-P, SP, PP is supported in part by the Robert
Welch Foundation and the NSF under grant number
PHY-9511632, JG is partially supported by AEN 98-0431 (CICYT), GC 1998SGR
(CIRIT),
SP is also supported by NSF grant PHY-9973543.

\appendix

\section{UV divergences in $g^2\f^4$}
Of the four terms in equation (\ref{uv}), the two damped by the Bose
distribution
$n_\b(|p|)$ are obviously UV convergent. The remaining two give rise to a
linear and
a logarithmic UV divergence (we remind the reader that the argument $|\t
p|$ of the
second Bose distribution $n_{1/\b}$ appearing in equation (\ref{uv}) is
independent
of $p_3$).

The term

$$-g^2 T \int \frac{d^3 p}{(2\p)^3}\frac{n_{1/\b}(|\t p|)}{ |p| \,4\p|\t p|}$$
is divergent in the region $p_3 \gg \sqrt{p_1^2 +p_2^2}$. In this region,
we obtain

$$-\frac{1}{2(2\p)^3}\frac{g^2 T}{\t} \log{\L}\int dp_{1,2} \, n_{1/\b}(|\t
p|)=-\frac{1}{2(2\p)^3}\frac{g^2}{\t^2} \log{\L} \int dx \, \frac{1}{e^{2\p
x}-1}
\,.$$
Notice that the $x$ integral is IR divergent. If we introduce an IR cutoff to
regulate this divergence, the substitution we use to extract the temperature
dependence of the expression renders this cutoff temperature dependent. We
ignore
this problem at this point, as we expect it to be an IR artifact that will
be cured
upon resummation of the appropriate diagrams.

The linear divergence arises as:

$$
-g^2 T \int \frac{d^3 p}{(2\p)^3} \frac{1}{2 |p| \,4\p|\t p|}=-\frac{1}{
4(2\p)^2}\frac{g^2 T}{\t}\L \,.
$$
As discussed above, this divergence contributes to the entropy of the
system, but
not to the internal energy. We therefore attribute it to the degeneracy of the
winding states as discussed in the text.

\section{IR divergences in $g^2\f^4$}

The only potential infrared divergence in the mean field expression
(\ref{ir}) for
the free energy stems from the term
$$\b V\langle
\int d^4x\,\f(x)*\f(x)*\f(x)*\f(x) \rangle.$$
The non-planar contribution to this term is obtained from equation
(\ref{eight}) by
replacing the bare by the dressed propagators:

\begin{equation}
-{g^2} T^2 \sum_{n,l} \int \frac{d^3p}{(2 \pi)^3}\, \frac{d^3k}{(2 \pi)^3}
\,\frac{e^{i\, p\, \t \, k}} {(\frac{4\pi^2 n^2}{\b^2} + p^2 + M^2+ g^2 \langle
\f^2 \rangle_{\t p})(\frac{4\pi^2 l^2}{\b^2} + k^2 + M^2+ g^2 \langle
\f^2 \rangle_{\t k})}  \,,
\end{equation}
where $\langle \f^2 \rangle$ is given to leading order by equation
(\ref{phi2}).
For small $p$ and small $k$, both terms in the denominator of the above
expression blow up and no infrared divergence can occur. For small $p$ and
large $k$
(or vice versa), we can ignore $\langle\f^2 \rangle_{\t k}$ and perform the
$d^3k$
integral as in (\ref{poisson}). Upon performing the sums, we obtain for
small $p$

\begin{eqnarray}
-{g^2} \int \frac{d^3p}{(2 \pi)^3} \, \frac{1 +
2 n_{\b}(|p|)}{2\sqrt{p^2 + M^2+ g^2 \langle\f^2 \rangle_{\t
p}}}\frac{2}{8\pi^2|\t
p|^2}\,.
\end{eqnarray}
As one factor of $|\t p|^{-1}$ is absorbed by the $\langle\f^2 \rangle_{\t
p}$ under
the square root, this expression is finite in the infrared.

\end{document}